# Band structure engineering using a moiré polar substrate

Xirui Wang[1], Cheng Xu[2], Samuel Aronson[1], Daniel Bennett[3], Nisarga Paul[1,4], Philip J.D. Crowley[7], Clément Collignon[1], Kenji Watanabe[5], Takashi Taniguchi[6], Raymond Ashoori[1], Efthimios Kaxiras[3,7], Yang Zhang[2,8], Pablo Jarillo-Herrero[1*], and Kenji Yasuda[1,9*]

[1]Department of Physics, Massachusetts Institute of Technology, Cambridge, Massachusetts 02139, USA.

[2]Department of Physics and Astronomy, University of Tennessee, Knoxville, Tennessee 37996, USA.

[3]John A. Paulson School of Engineering and Applied Sciences, Harvard University, Cambridge, Massachusetts 02138, USA.

[4]Kavli Institute of Theoretical Physics, University of California, Santa Barbara, Santa Barbara, California 93106, USA.

[5]Research Center for Electronic and Optical Materials, National Institute for Materials Science, 1-1 Namiki, Tsukuba 305-0044, Japan.

[6]Research Center for Materials Nanoarchitectonics, National Institute for Materials Science, 1-1 Namiki, Tsukuba 305-0044, Japan.

[7]Department of Physics, Harvard University, Cambridge, Massachusetts 02138, USA.

[8]Min H. Kao Department of Electrical Engineering and Computer Science, University of Tennessee, Knoxville, Tennessee 37996, USA.

[9]School of Applied and Engineering Physics, Cornell University, Ithaca, New York 14850, USA.

*Correspondence to: pjarillo@mit.edu, kenji.yasuda@cornell.edu

**Abstract:**

Applying long wavelength periodic potentials on quantum materials has recently been demonstrated to be a promising pathway for engineering novel quantum phases of matter. Here, we utilize twisted bilayer boron nitride (BN) as a moiré substrate for band structure engineering. Small-angle-twisted bilayer BN is endowed with periodically arranged up and down polar domains, which imprints a periodic electrostatic potential on a target two-dimensional (2D) material placed on top. As a proof of concept, we use Bernal bilayer graphene as the target material. The resulting modulation of the band structure appears as superlattice resistance peaks, tunable by varying the twist angle, and Hofstadter butterfly physics under a magnetic field. Additionally, we demonstrate the tunability of the moiré potential by altering the dielectric thickness underneath the twisted BN. Finally, we find that near-60°-twisted bilayer BN provides a unique platform for studying the moiré structural effect without the contribution from electrostatic moiré potentials. Tunable moiré polar substrates may serve as versatile platforms to engineer the electronic, optical, and mechanical properties of 2D materials and van der Waals heterostructures.

**Main text:**

The emerging concept of quantum metamaterials enables us to engineer electronic structures and physical properties that do not exist in natural crystals[1]. This is most strikingly exemplified by moiré materials, where two-dimensional (2D) materials are stacked at controlled angles. For instance, twisted bilayer graphene and transition metal dichalcogenide moiré bilayers exhibit topological and strongly-correlated phases of matter, absent in their constituent layers[2–11]. Despite the prosperous findings in these moiré systems, the method of creating a moiré potential by twisting the target material itself poses limitations in the choice of materials, moiré periodicity, and potential strength. This is mainly due to the requirement of the proximity of two atomic layers with similar or identical lattice constants.

A remote tunable superlattice potential isolated from a target layer can overcome these constraints. Past research has etched periodic holes in gate dielectrics or gate electrodes via electron beam lithography or focused ion beam milling, and introduced an electrostatic periodic potential in the target layers via gating[12–16]. The effect of a periodic potential on graphene was observed as superlattice resistance peaks due to band folding and Hofstadter physics[12–16]. Despite the versatility in shape and the tunability in the potential strength, this method suffers from two limitations. First, the pitch resolution is technically restricted when the pattern is defined by electron beam lithography or focused ion milling, being difficult to achieve below 16 nm[14], because of the secondary electrons or limitation in beam size. Second, as each pitch is written independently through lithography or focused ion beam, the unavoidable non-identicality among moiré sites works as disorder in the moiré potential, leading to broadened superlattice peaks and potentially smearing out detailed features. A natural way to overcome this issue is to create moiré potentials via twisted van der Waals (vdW) materials, which can achieve arbitrary moiré periodicity by controlling the twist angle, and highly periodic moiré superlattices formed from identical unit cells due to their crystalline nature[17–19].

We engineer such a moiré substrate using a twisted 2D insulator — twisted bilayer boron nitride (BN). Hexagonal BN is one of the most important constituents of vdW heterostructures, which is primarily used as an atomically-flat substrate and gate dielectric. Recently, it has been demonstrated that out-of-plane ferroelectricity can be obtained by stacking two monolayers of BN in parallel[20–22]. Furthermore, a moiré polar pattern is generated by introducing a small angle between the monolayers (Fig. 1b), which is characterized by upward and downward polarizations periodically arranged in the lateral direction[20–23]. Using twisted bilayer BN as a substrate, we can apply a moiré electrostatic potential to arbitrary 2D materials[24]. Here, we use bilayer graphene as the target material, illustrating the resultant band structure modification and the tunability of the twisted BN moiré substrates.

A monolayer BN consists of boron (B) and nitrogen (N) atoms alternately arranged at honeycomb lattice sites. When two BN monolayers are twisted at a small angle away from parallel stacking, a moiré pattern consisting of AA, AB, and BA stacking regions is formed as illustrated in Fig. 1b. In AA stacking, the top and bottom layers directly overlap, having no net polarization. In AB (BA) stacking, meanwhile, B (N) sits on top of the N (B) atom, forming an electric dipole and giving rise to downward (upward) polarization (Fig. 1a)[21,22]. We consider such twisted bilayer BN as forming a moiré polar substrate, referring to the fact that the moiré potential comes from the electrostatic potential in these staggered up and down polar domains. Fig. 1c illustrates the schematic of our device, where we placed a target layer on top of the twisted BN moiré substrate, controlled with a top metal gate and a bottom graphite gate. In this device geometry, we expect the modulation of the potential $U$ on the target layer yielding a peak magnitude of around 29 mV when the moiré wavelength is 18.5 nm, and the target layer is bilayer graphene (Fig. 1d, see details of the simulation in Methods). Fig. 1e displays the theoretical calculation of the band structure. The superlattice moiré potential leads to the band folding into a mini-Brillouin zone, with density of state (DOS) minima at the band edges (see details of the calculation in Section 1 of Supplementary Materials).

**Band structure modulation in bilayer graphene:**

We fabricated a series of bilayer graphene/twisted bilayer BN devices (A1-A5), aiming at different twist angles between 0.4° and 1.3°. We first made the top stack by sequentially picking up the top BN, bilayer graphene, and twisted bilayer BN using the tear-and-stack technique. We intentionally misaligned the bilayer graphene with top and bottom BN layers to avoid the formation of long-periodicity moiré patterns by the BN to graphene alignment. Subsequently, we performed piezoresponse force microscopy (PFM) to identify regions with uniform moiré patterns. Finally, we placed the top stack onto the bottom graphite, etched and contacted the device. The detailed parameters of the devices are summarized in Extended Data Table 1, and the optical microscope images are shown in Extended Data Fig. 1. Fig. 2a plots the longitudinal sheet resistivity $\rho_{xx}$ as a function of carrier density $n$, calibrated from Landau fan diagram measurements. We observe satellite resistance peaks symmetrically located around the charge neutrality point (CNP) at different carrier densities for A1-A5. According to the band structure calculation in Fig. 1e, we expect DOS minima, *i.e.*, resistance peaks, at the filling of one moiré band on both the electron and hole sides, corresponding to $n_0 = 4/A$, where $A = \sqrt{3}/2\, a^2$ is the moiré unit cell area, and the prefactor 4 corresponds to the spin and valley degeneracy. The moiré wavelength of device A2 calculated from the satellite resistance peak positions is around 13.4 nm, which is consistent with the one obtained from the PFM image, 11.9 nm (Fig. 2c). This agreement confirms that the satellite resistance peaks originate from the band structure modulation induced by the twisted BN moiré substrate.

Our method can produce arbitrary moiré periodicity by tuning the twist angle, which is in contrast to the moiré periodicity formed by BN to graphene alignment. The latter cannot be longer than 14 nm because of the finite lattice mismatch. We also note that when the moiré potential was induced by aligning a piece of BN to graphene, the hole side typically featured a more prominent resistance peak than the electron side[25–27], while in our devices, the heights of resistance peaks on both sides are comparable, in line with the symmetric DOS minima on the electron and hole sides in Fig. 1e and Extended Data Fig. 2.

The Hall resistivity $\rho_{yx}$ under a small magnetic field ($B = 0.2$ T) as a function of $n$ shows sign reversals at the satellite resistance peak positions as well as between CNP and satellite resistance peak positions (Fig. 2b). The former can be attributed to the moiré band edges, and the latter to the Van Hove singularities (VHSs). We note that both the satellite resistance peaks in $\rho_{xx}$ and sign reversals in $\rho_{yx}$ become stronger when the moiré wavelength increases. We can qualitatively understand this trend as the following: since the wavevector is inversely proportional to the moiré wavelength, as the moiré wavelength increases, the kinetic energy at the mini-Brillouin zone boundary decreases. Therefore, the relative effect of the electrostatic potential is enhanced, leading to a stronger band reconstruction.

We further investigate the magnetic field response by increasing the magnetic field and plotting the longitudinal conductance $\sigma_{xx}$ versus filling factor $n/n_0$ and the normalized magnetic flux $\phi/\phi_0$ in Fig. 2d. We observe Landau fan features emanating from $n/n_0 = +4$ and $-4$, respectively, indicating the formation of moiré mini bands[25–27]. We also observe Brown-Zak oscillations, corresponding to enhanced conductivity when the cyclotron orbits are commensurate with the moiré wavelengths at $\phi/\phi_0 = $ ½, ⅓, ¼, and so on[28]. In the Landau fan diagram of transverse conductance $\sigma_{xy}$, the sign reversals diminish at a critical field of around 0.7 T on the hole side and 0.3 T on the electron side due to the magnetic breakdown (Fig. 2e). The difference in the critical magnetic fields points to the electron-hole asymmetry in the system. We can estimate the strength of the moiré potential based on the VHS location, which is expected to happen at a smaller filling factor as the moiré potential becomes larger (see calculation details in Extended data Fig. 2 and Section 1 of Supplementary Materials). From the VHS location of $n/n_0 \approx -1.9$ on the hole side, we quantify the peak potential magnitude to be 105 mV, which is of the same order of magnitude as (albeit bigger than) our simulated moiré potential strength. As shown in the temperature dependence in Extended Data Fig. 3, the satellite resistance peaks persist up to 100 K, which is consistent with the moiré potential energy scale. Our transport measurements, together with PFM characterization and theoretical calculations, demonstrate that twisted bilayer BN can work as a substrate to engineer moiré superlattices with different wavelengths and induce band structure modifications to 2D materials. The satellite resistance peaks, and the sign reversals of the Hall resistance are also observable in monolayer graphene on twisted BN (Extended Data Fig. 4), demonstrating the versatility of our moiré band engineering method.

**Tuning moiré potential by inserting extra dielectric BN:**

Next, we demonstrate the tunability of the strength of the moiré potential by inserting extra dielectric BN underneath the twisted bilayer BN. When a thick BN is inserted between the bottom gate and a ferroelectric bilayer BN (in this paper, "thick BN" stands for 5 to 80 nm), the doping induced in the target layer is suppressed due to the larger distance between the target layer and the gate electrode[21]. In Fig. 3a, we simulated the superlattice potential induced by the twisted bilayer BN at different total bottom BN thicknesses, denoted as $d$. The peak potential $U_{peak}$ is plotted as a function of $d$, decreasing from 29 mV in the case without bottom thick BN to 11 mV at $d \geq 10$ nm (details of the simulation are shown in Extended Data Fig. 5). Fig. 3b presents a comparison between two devices with a similar moiré wavelength: device A3 without bottom thick BN and device C1 with bottom thick BN ($d$ = 15.1 nm). The heights of satellite resistance peaks in device C1 are less than one third of those in device A3, consistent with the smaller magnitude of the moiré potential.

As the magnetic field is turned on, we observe a similar yet weaker Hofstadter spectrum in device C1, with Landau levels emerging from moiré band edges, as well as Brown-Zak oscillations (Fig. 3c). The sign reversals in $\sigma_{xy}$ diminish at around 0.3 T on the hole side, and no sign reversal is observed on the electron side except for a weakly reduced Hall conductance. The VHS is located at $n/n_0 \approx -3.0$ on the hole side, from which we extracted the peak potential to be 41 mV, around 40% of the value we extracted in device A3. The shifts in VHS locations, the lower satellite peak heights, weaker Hofstadter spectrum, and smaller critical field of Hall sign reversals, are all consistent with the weaker moiré potential in device C1 as compared to device A3. Additional data of more devices with inserted thick bottom BN are shown in Extended Data Fig. 6.

**Moiré structural effect from near-60°-twisted BN:**

So far, we have only considered the electrostatic effect from the polar domains in the twisted bilayer BN. The structural lattice deformation effect has also been considered important in moiré heterostructures and band structure modulation[18,29–37]. Experimentally, however, it has been challenging to distinguish it from the electrostatic effect, as these two effects typically coexist in moiré superlattices formed by 2D materials. For example, in near-0°-twisted BN, the dominant moiré electrostatic potential makes it difficult to isolate the moiré structural effect. Here, we show that near-60°-twisted BN is an ideal platform for investigating the moiré structural effect independently. Different from the near-0°-twisted case, near-60°-twisted bilayer BN consists of AA', AB', and BA' stacking arrangements (Fig. 4b). Here, ' means one layer is rotated by 60° with respect to the other. The AA' stacking has each B atom stacked on top of an N atom or vice versa, and the AB' (BA') stacking has N (B) stacked on top of N (B) atoms (Fig. 4a). The interlayer distance ($d$) between the two monolayers of BN varies depending on the stacking order, with spatial modulation as large as 7% according to the density functional theory (DFT) calculations (Fig.

4f, see details of the calculation in Section 2 of Supplementary Materials). Importantly, all of these stacking orders have an inversion center, *i.e.*, the top and bottom layers are connected by the inversion operation. This also holds true for low-symmetry stacking orders other than AA', AB', and BA'. Hence, the polar domains and associated electrostatic potential are absent in near-60°-twisted BN, leaving only the moiré structural effect (Fig. 4a).

To study such structural effect on the electronic band structure, we prepared a device consisting of a bilayer graphene on top of a 60.90°-twisted bilayer BN, encapsulated by top and bottom thick BNs (Fig. 4c). We observed small satellite resistance peaks in $\rho_{xx}$ plotted as a function of *n*, corresponding to the full filling of the moiré unit cell (Fig. 4d,e). Since there is no electrostatic potential modulation from the twisted BN, the resistance peaks are attributed to the periodic lattice deformation in the BN bilayer. Additional data from devices with near-60°-twisted BN are included in Extended Data Fig. 7.

Although we do not have a unified understanding of how the moiré structure modulation affects the band structure, we propose one way to understand it through the out-of-plane corrugation effect. We assume that the bottom layer in the bilayer graphene follows the corrugated twisted BN profile, while the top layer which is in contact with the flat top BN only partially follows the profile modulation. In this case, the interlayer distance, and consequently the tunneling strength, between the top and bottom layers of graphene is modulated by the moiré periodicity, giving rise to band folding. In the real situation, both the out-of-plane corrugation and the in-plane strain can contribute to the band structure modulation in an intricate way[29]. The exact mechanism of the moiré structural effect calls for future experimental and theoretical studies.

**Conclusions:**

In conclusion, we demonstrated that twisted bilayer BN can serve as a versatile moiré substrate for band structure engineering with tunable wavelength and potential strength. Our moiré polar substrate design principle is applicable not only to BN, but also to other bipartite 2D insulators or semiconductors, such as transition metal dichalcogenides, which have different magnitudes of electrostatic potential and spin-orbit interaction[38,39]. The application of the moiré substrates on various 2D materials may lead to the observation of exotic physics related to correlations and topology in the future[24,40–45]. Moreover, this lithographically free method for introducing highly periodic electrostatic potentials at the nanoscale may find applications in other fields of nanotechnology beyond 2D materials[46–48].

**Methods:**

Device fabrication:

BN and graphite crystals were exfoliated onto $SiO_2$ (285 nm-295 nm)/Si substrates. For monolayer BN, we used $SiO_2$ (90 nm)/Si substrates. The thickness of thick BN flakes (> 5 nm) was acquired by atomic force microscopy (AFM). Monolayer BN, graphene and bilayer graphene were identified by optical contrast. We first prepared the bottom stacks. For devices A1-A5 and B1, we exfoliated graphite for bottom gates on $SiO_2$ (285 nm)/Si substrates with pre-patterned markers, followed by heat cleaning in an atmosphere of Ar (40 sccm) and $H_2$ (20 sccm) gases at 350 °C for more than 12 hours to remove the tape residues. For devices C1-C5, D1, and D2, a thick BN (5~20 nm) and graphite were sequentially picked up by poly(bisphenol A carbonate) (PC)-film-covered Polydimethylsiloxane (PDMS) stamp on a glass slide, and then released to a $SiO_2$/Si substrate at 170 °C. We removed the trapped bubbles by moving the PC stamp and the stack up and down slowly a few times at 170 °C. After the PC was dissolved, the bottom stacks were heat cleaned in an atmosphere of Ar (40 sccm) and $H_2$ (20 sccm) gases at 350 °C for 3~12 hours. After that, contact mode AFM was performed at a deflection voltage of 0.1~0.2 V for cleaning the PMMA residues.

Each top stack was made by the sequential pickup of top BN, bilayer/monolayer graphene, and twisted bilayer BN. Twisted bilayer BN was obtained by the sequential pickup of monolayer BN flake at room temperature with the tear-and-stack method described in Refs 50 and 51. Then the stack on PC was scanned by PFM to search for regions with periodic moiré patterns. The whole stack was then released to the bottom graphite or bottom BN/graphite stack at 170 °C. The stack was identified with an optical microscope and AFM for bubble-free regions. Then the stack was etched into a Hall bar shape by reactive ion etching, leaving the bubble-free regions with periodic moiré patterns remained. All the contacts and top gates were deposited with Cr/Au with a thermal evaporator.

Piezoresponse force microscopy measurements:

We performed lateral PFM measurements on the top stack containing twisted bilayer BN. The PFM measurements were performed with Asylum Research Cypher S atomic force microscope at room temperature. We used AC240TM-R3 tip with a force constant of around 1.5 N m$^{-1}$ and a contact resonance frequency of around 600 kHz with the applied AC bias voltage of 2 V. The contact strength was set to be lower than 30 nN to avoid unintentional damage to the flake and twist angle relaxation.

Transport measurements:

The devices were bonded by aluminum wire. The four-probe measurements were done using lock-in amplifiers (SRS: SR830 and SR860), a current preamplifier (DL: Model 1211) and voltage preamplifiers (SRS: SR560) at the frequency of 17~35 Hz. The gate voltages were applied by source meters (Keithley: Model 2400 and 2450). Devices were measured in a He-3 cryostat (Janis research) or in a variable temperature insert (Cryogenics).

Electrostatic simulation:

We used COMSOL Multiphysics® to simulate the electrostatic potential in bilayer graphene.

Initial condition and boundary condition: First, we built the geometries composed of the bilayer graphene, twisted bilayer BN, (bottom thick BN,) and bottom graphite in shapes of rectangular cuboids (Extended Data Fig. 5). The device lateral size was set to be 80 nm × 80 nm, and the thicknesses were assigned according to the device structures that we study. We set the moiré wavelength of twisted bilayer BN to be 18.5 nm. For the initial condition, we assume sinusoidal 2D charge density distribution on either side of the twisted bilayer BN, taking the form $\rho = 2\rho_0\left(\sin(kx + \sqrt{3}ky) - \sin 2kx + \sin(kx - \sqrt{3}ky)\right)/3\sqrt{3}$, where $k = 1/a$. The peak value of charge density is taken to be $\rho_0 = \epsilon_0 \epsilon_{BN} V_p / d_B$, where $\epsilon_0$ is the vacuum permittivity, $V_p = 0.109$ V is the interlayer potential, and $d_B$ is the interlayer distance between bilayer BN[21,38]. For the boundary condition, we set the bilayer graphene and bottom graphite to be grounded, i.e., chemical potential $V_0 = 0$.

Quantum capacitance treatment: Since bilayer graphene is not a perfect metal and has finite DOS, we need to consider the effect from its quantum capacitance. After the first round of simulation as described above, we acquired the electric field distribution close to the bilayer graphene, and calculated the doping induced by the electric field using Gauss's law: $n = \epsilon_0 \epsilon_{BN} E$, where $\epsilon_{BN}$ is taken to be 3. Then we calculated the chemical potential shift $V_1$ in bilayer graphene due to this doping effect. We then set the new boundary condition at bilayer graphene to be $V_2 = 1/2(V_0 + V_1)$, and performed the simulation again. We performed the above steps in an iterative way until the electric potential in bilayer graphene converges, where the system reaches the electrochemical equilibrium. This means that the electrostatic potential shift due to geometric capacitance and the chemical potential shift due to quantum capacitance are both considered, and the potential profile is self-consistent. To calculate $V_1$, we first acquired Fermi wavevector from $k_F = \sqrt{\pi n}$, considering the valley and spin degeneracy in bilayer graphene, and then derived $V_1$ from $k_F$ using the dispersion relation calculated from equation (1) in Supplementary Materials.

**Acknowledgements:**

We thank A. Reddy for help in fabrication, and J. Cano, S. A. A. Ghorashi, and M. Koshino for fruitful discussions. This research was supported by the Center for the Advancement of Topological Semimetals, an Energy Frontier Research Center funded by the U.S. Department of Energy Office of Science, through the Ames Laboratory under contract DE-AC02-07CH11358 (measurements and data analysis), the MIT/Microsystems Technology Laboratories Samsung Semiconductor Research Fund, the Gordon and Betty Moore Foundation's EPiQS Initiative through grant GBMF9463, and the Ramon Areces Foundation. This work was performed in part at the Harvard University Center for Nanoscale Systems (CNS), a member of the National Nanotechnology Coordinated Infrastructure Network (NNCI), which is supported by the National Science Foundation under NSF ECCS award No. 1541959. C. X. and Y. Z. are supported by the start-up fund at University of Tennessee Knoxville, and the National Science Foundation Materials Research Science and Engineering Center program through the UT Knoxville Center for Advanced Materials and Manufacturing (Grant No. DMR2309083). S.A. is partially supported by the NSF Graduate Research Fellowship Program via grant no. 1122374. D.B. and E.K. acknowledge the US Army Research Office (ARO) MURI project under grant No. W911NF-21-0147 and from the Simons Foundation award No. 896626. N. P. acknowledges the Kavli Institute for Theoretical Physics (KITP) graduate fellowship. K.W. and T.T. acknowledge support from the JSPS KAKENHI (Grant Numbers 21H05233 and 23H02052) and World Premier International Research Center Initiative (WPI), MEXT, Japan.

Figures:

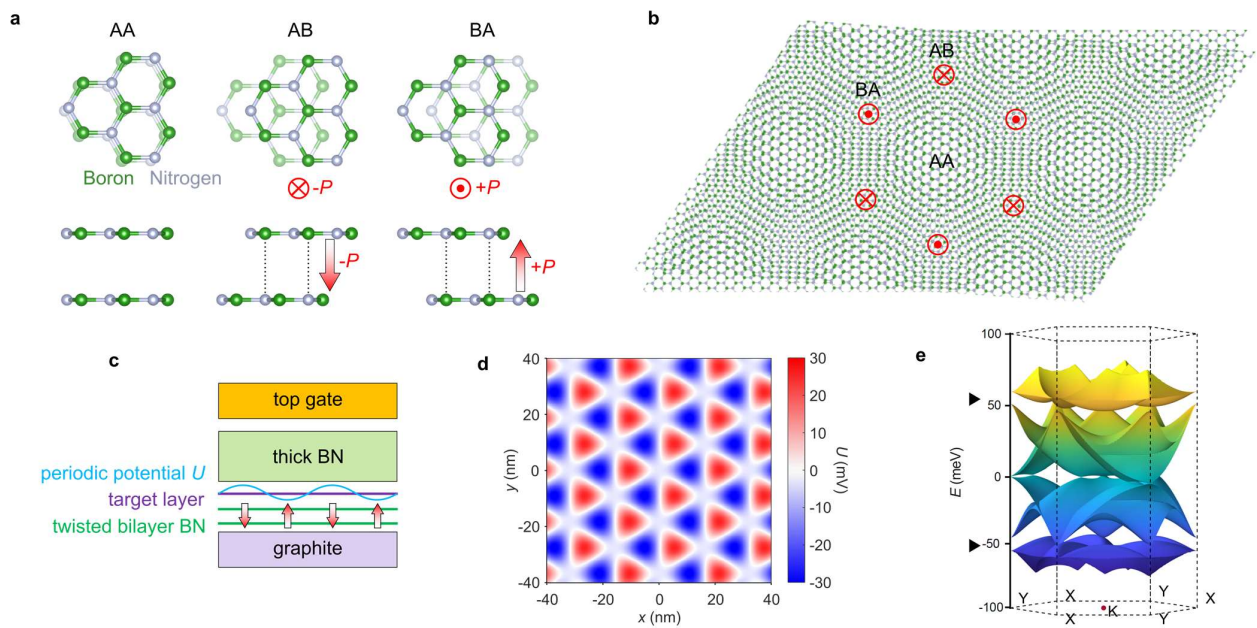

**Fig. 1 | Twisted bilayer BN as a moiré polar substrate. a,** Three high-symmetry stacking orders that exist in near-0°-twisted bilayer BN[49]. N and B atoms are shown in silver and green, respectively. In AA stacking (left), top and bottom atoms align right on top of each other, and there is no out-of-plane polarization. In AB (middle) and BA (right) stacking, the vertical alignment of N and B atoms creates an out-of-plane electric dipole, leading to downward (upward) polarization in AB (BA) stacking. **b,** Schematic of twisted bilayer BN, where moiré patterns form with AB (downward polarization), BA (upward polarization), and AA local stacking arrangements[49]. **c,** Device schematic for using twisted BN as a moiré polar substrate. Target layer sits on top of twisted bilayer BN, feeling its periodic moiré potential $U$. The target layer is encapsulated by a thick BN on top, and electrically gated with a metal top gate and a graphite bottom gate. **d,** Electrostatic simulation of potential strength $U$ imposed on bilayer graphene in the device structure shown in **c**, and the moiré wavelength $a$ is taken to be 18.5 nm. **e,** Band structure calculation of bilayer graphene under the electrostatic moiré potential in **d**. Arrows mark the locations of DOS minima in the conduction band and valence band due to band folding induced by the superlattice potential.

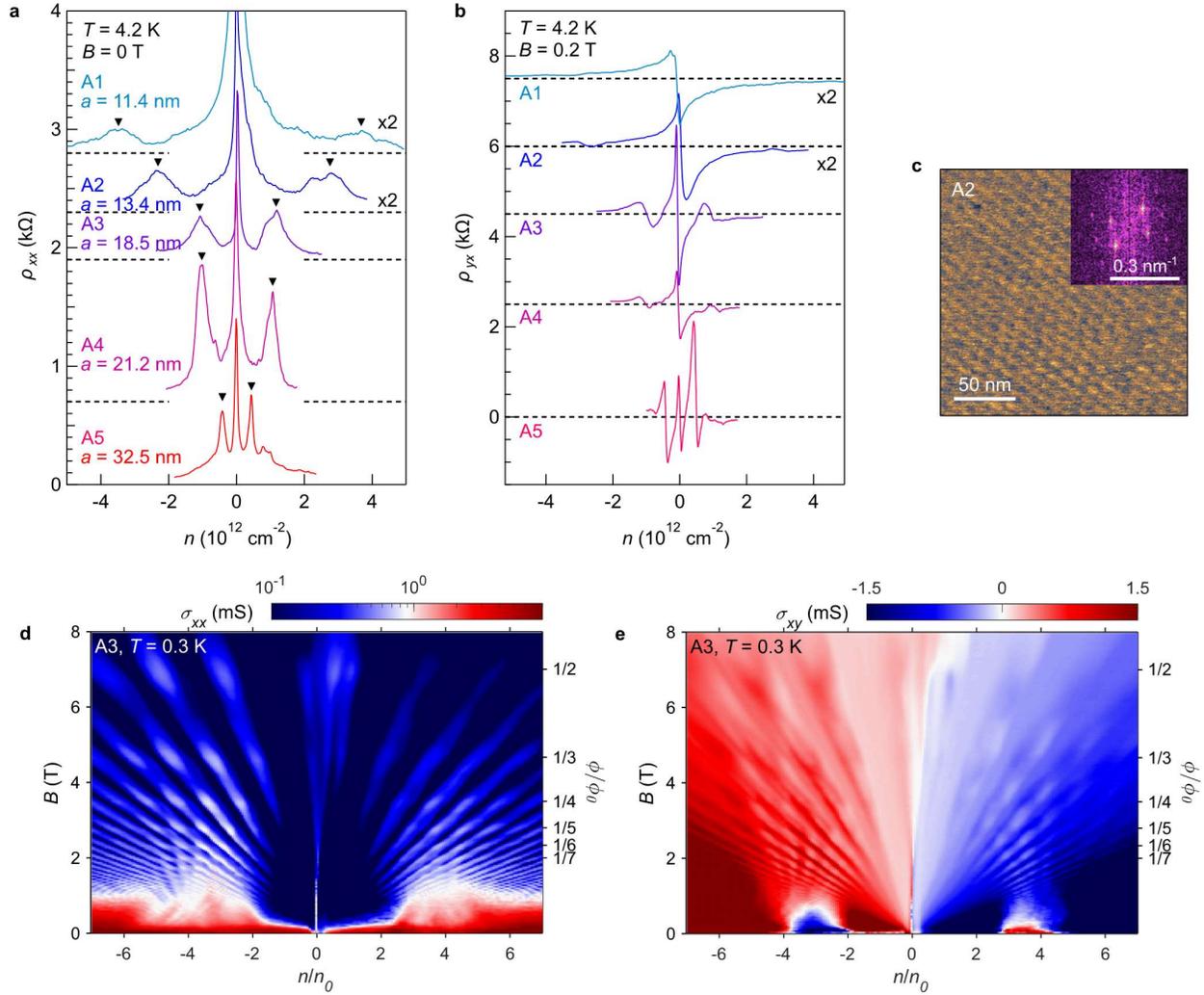

**Fig. 2 | Band structure modulation of bilayer graphene on twisted bilayer BN moiré polar substrate. a,** Longitudinal resistivity $\rho_{xx}$ as a function of carrier density $n$ in devices A1-A5, measured at 4.2 K. Satellite resistance peaks symmetrically located around CNP are observed at different carrier densities that correspond to different moiré wavelengths. Arrows indicate the positions of the satellite resistance peaks. **b,** Hall resistivity $\rho_{yx}$ as a function of $n$ in devices A1-A5 at $B = 0.2$ T, $T = 4.2$ K. **c,** PFM image of twisted BN before making into device A2. Scale bar: 50 nm. Inset: Fourier transform image. Scale bar: 0.3 nm$^{-1}$. **d,** Longitudinal conductance $\sigma_{xx}$ as a function of filling factor $n/n_0$ and magnetic field $B$ (right y axis: normalized magnetic flux $\phi/\phi_0$), measured at 0.3 K in device A3. $\phi = BA$, the magnetic flux per moiré unit cell, and flux quantum $\phi_0 = h/e$. **e,** Transverse conductance $\sigma_{xy}$ as a function of $n/n_0$ and $B$ (right y axis: $\phi/\phi_0$), measured at 0.3 K in device A3.

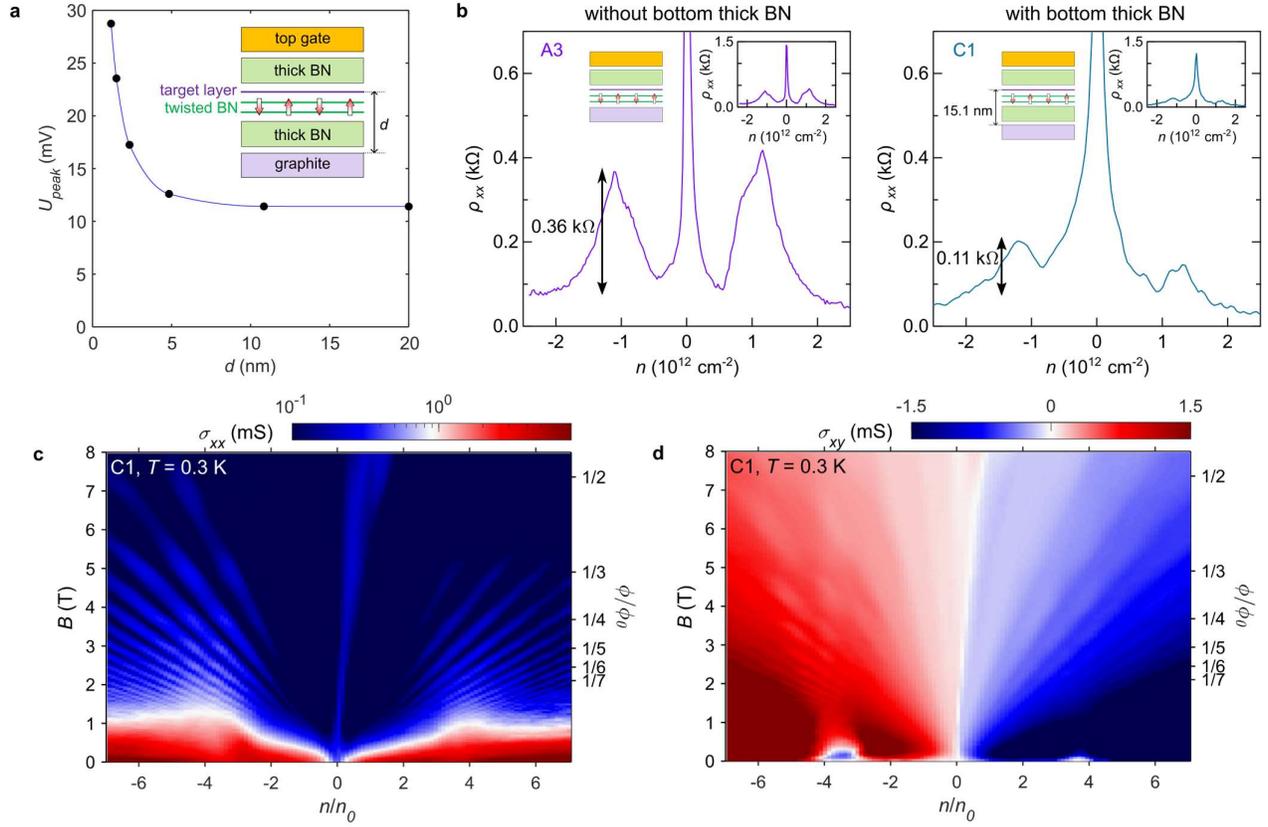

**Fig. 3 | Tuning moiré polar potential by changing dielectric thickness. a,** Electrostatic simulation of the moiré potential peak magnitude $U_{peak}$ as a function of $d$. Inset: Device schematic. A thick BN layer is inserted between the twisted bilayer BN and the bottom graphite. $d$: total bottom BN thickness. **b,** $\rho_{xx}$ as a function of $n$ in devices A3 (without bottom thick BN) and C1 (with bottom thick BN), measured at 4.2 K. Insets: Device schematics (left) and large-scale plots (right). **c,** $\sigma_{xx}$ as a function of $n/n_0$ and $B$ (right y axis: $\phi/\phi_0$), measured at 0.3 K in device C1. **d,** $\sigma_{xy}$ as a function of $n/n_0$ and $B$ (right y axis: $\phi/\phi_0$), measured at 0.3 K in device C1.

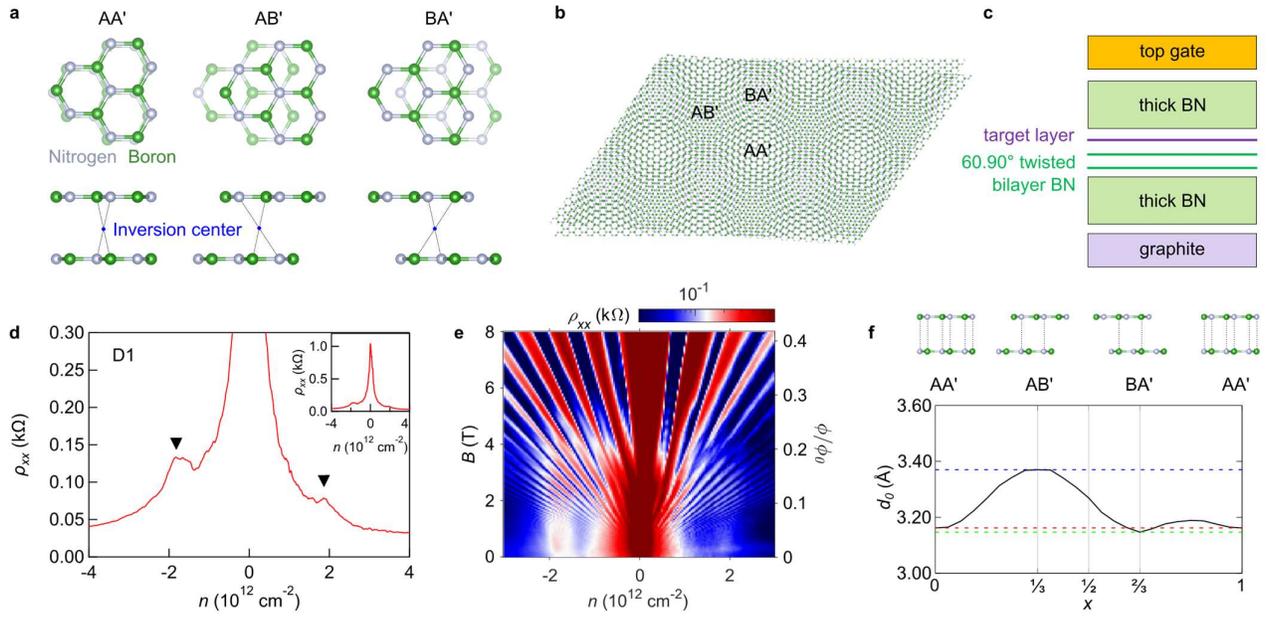

**Fig. 4 | Moiré potential in near-60°-twisted bilayer BN. a,** Three stacking orders that exist in near-60°-twisted bilayer BN[49]. In AA' stacking (left), B and N atoms align on top of each other alternatively. Here, ' means one layer is rotated by 60° with respect to the other. In AB' stacking(middle), N sits on top of N. In BA' stacking(right), B sits on top of B. All of these stacking arrangements have an inversion center, and therefore no net polarization is allowed. **b,** Schematic of near-60°-twisted bilayer BN, where moiré patterns form with AA', AB', and BA' local stacking arrangements[49]. **c,** Device schematic of D1. The target layer sits on top of a near-60°-twisted bilayer BN. The stack is encapsulated between the top and bottom BN and controlled by a top metal gate and a bottom graphite gate. **d,** $\rho_{xx}$ as a function of $n$ in device D1 with near-60°-twisted bilayer BN, measured at 4.2 K. Arrows indicate the positions of the satellite resistance peaks. Inset: Large-scale plot. **e,** $\rho_{xx}$ as a function of $n$ and $B$ (right $y$ axis: $\phi/\phi_0$), measured at 0.3 K in device D1. **f,** DFT calculation of the interlayer distance $d_0$ between the two near-60°-twisted monolayer BNs at different moiré sites. $x$: fractions of the moiré unit cell.

**Extended Data Table 1 | List of devices**

| Device | $d_T$ (nm)[#] | $d$ (nm) | Target layer | Bottom BN | $a$ (nm)[*] | Angle[*] |
|---|---|---|---|---|---|---|
| A1 | 55 | N.A. | bilayer graphene | absent | 11.4 | 1.25° |
| A2 | 42 | N.A. | bilayer graphene | absent | 13.4 | 1.07° |
| A3 | 48 | N.A. | bilayer graphene | absent | 18.5 | 0.78° |
| A4 | 55 | N.A. | bilayer graphene | absent | 21.2 | 0.68° |
| A5 | 33 | N.A. | bilayer graphene | absent | 32.5 | 0.44° |
| B1 | 50 | N.A. | monolayer graphene | absent | 11.9 | 1.20° |
| C1 | 28 | 15.1 | bilayer graphene | exist | 18.0 | 0.80° |
| C2 | 42 | 16.2 | bilayer graphene | exist | 14.4 | 1.00° |
| C3 | 28 | 17.7 | bilayer graphene | exist | 20.0 | 0.71° |
| C4 | 74 | 8.5 | bilayer graphene | exist | 34.6 | 0.41° |
| D1 | 43 | 8.9 | bilayer graphene | exist | 15.8 | 60.90° |
| D2 | 38 | 16.0 | bilayer graphene | exist | 11.2 | 61.28° |

[#]: Top BN thickness.
[*]: $a$ and angles are both calculated based on the electrical transport measurement features.

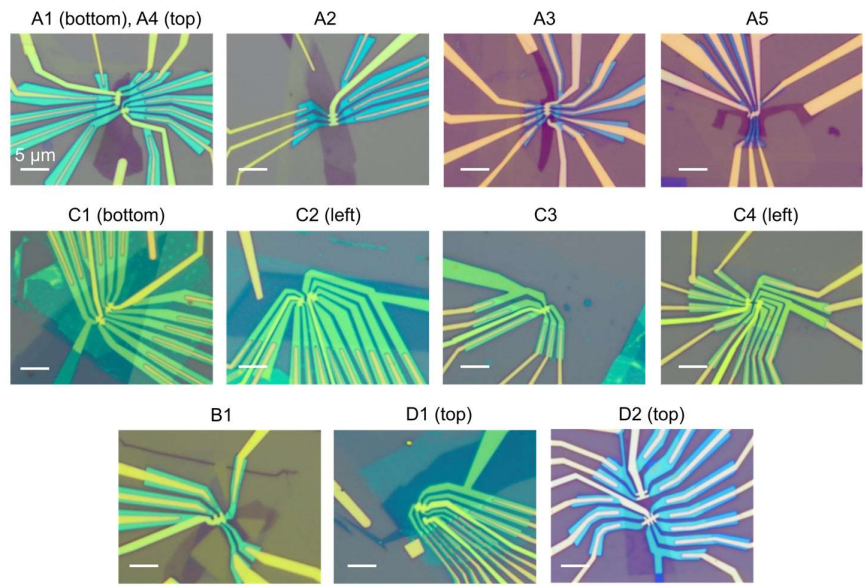

**Extended Data Fig. 1 | Optical images of devices used in this study.** The device used in this study is indicated in the parentheses for the images with two devices. Scale bar: 5 μm.

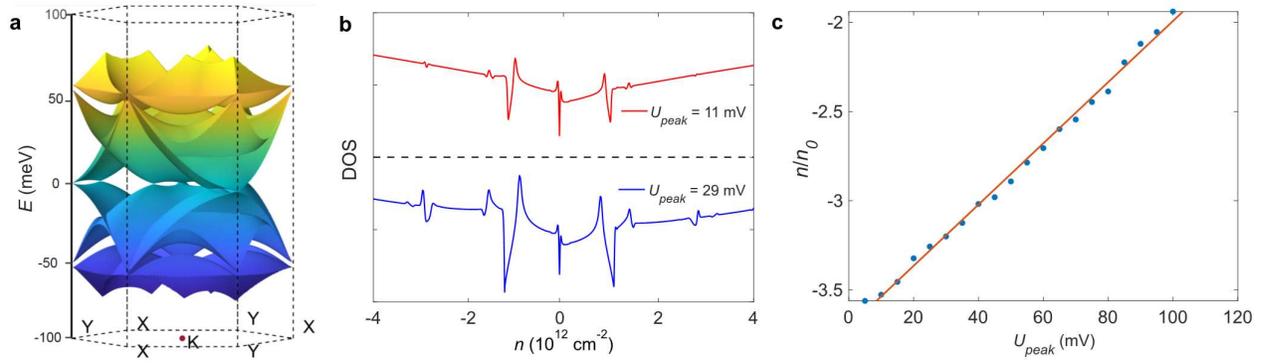

**Extended Data Fig. 2 | Band structure calculations. a,** Band structure of bilayer graphene at $a$ = 18.5 nm, and $U_{peak}$ = 11 mV. **b,** Calculated DOS as a function of $n$ in bilayer graphene under different potential strengths at $a$ = 18.5 nm. The larger moiré potential leads to smaller DOS at band edges. **c,** Expected VHS location ($n/n_0$) as a function of $U_{peak}$ at $a$ = 18.5 nm. The dots are the data points where the calculations were performed, and the solid line is the linear fitting.

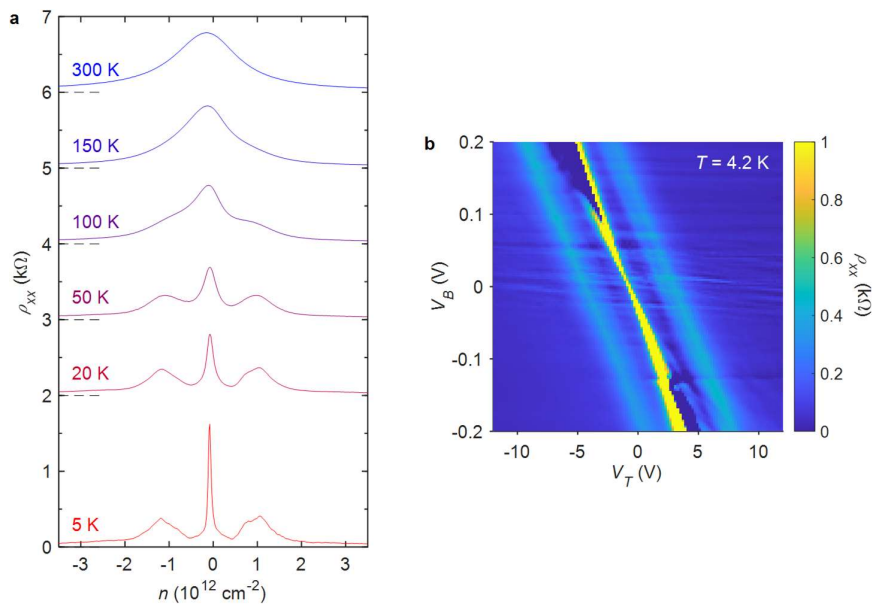

**Extended Data Fig. 3 | Additional data of device A3. a,** Temperature dependence of $\rho_{xx}$ as a function of $n$. **b,** Dual-gate scan of $\rho_{xx}$ measured at 4.2 K.

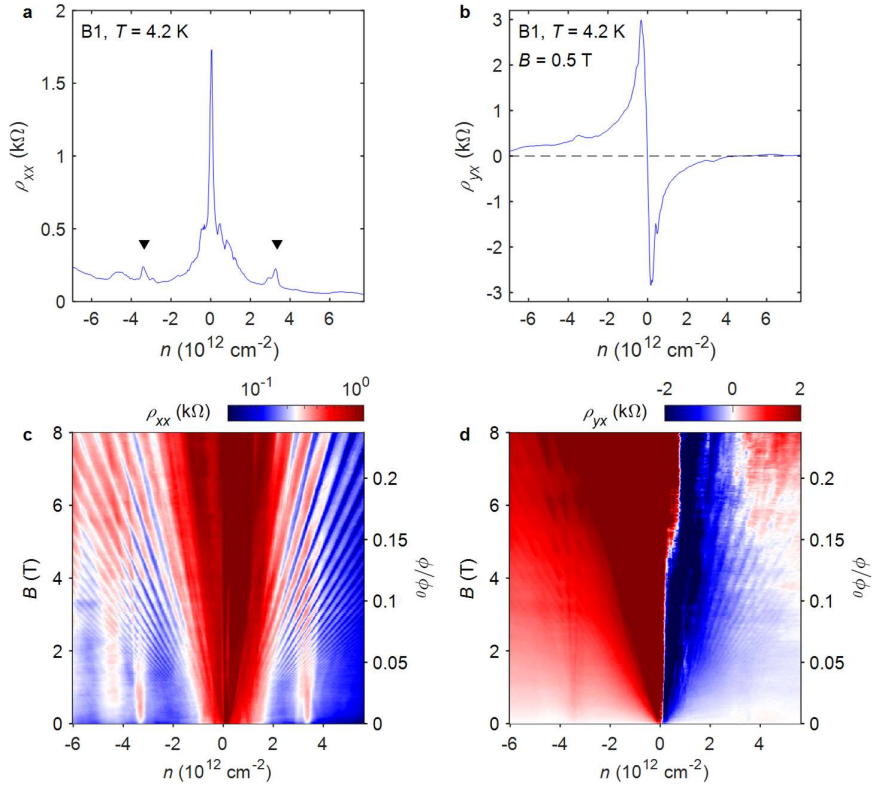

**Extended Data Fig. 4 | Band structure modulation of monolayer graphene on twisted bilayer BN. a,** $\rho_{xx}$ as a function of $n$. Arrows indicate the positions of the satellite resistance peaks. **b,** $\rho_{yx}$ as a function of $n$ at $B = 0.5$ T. **c,** $\rho_{xx}$ as a function of $n$ and $B$ (right $y$ axis: $\phi/\phi_0$). **d,** $\rho_{yx}$ as a function of $n$ and $B$ (right $y$ axis: $\phi/\phi_0$). All the data were taken in device B1, at $T = 4.2$ K.

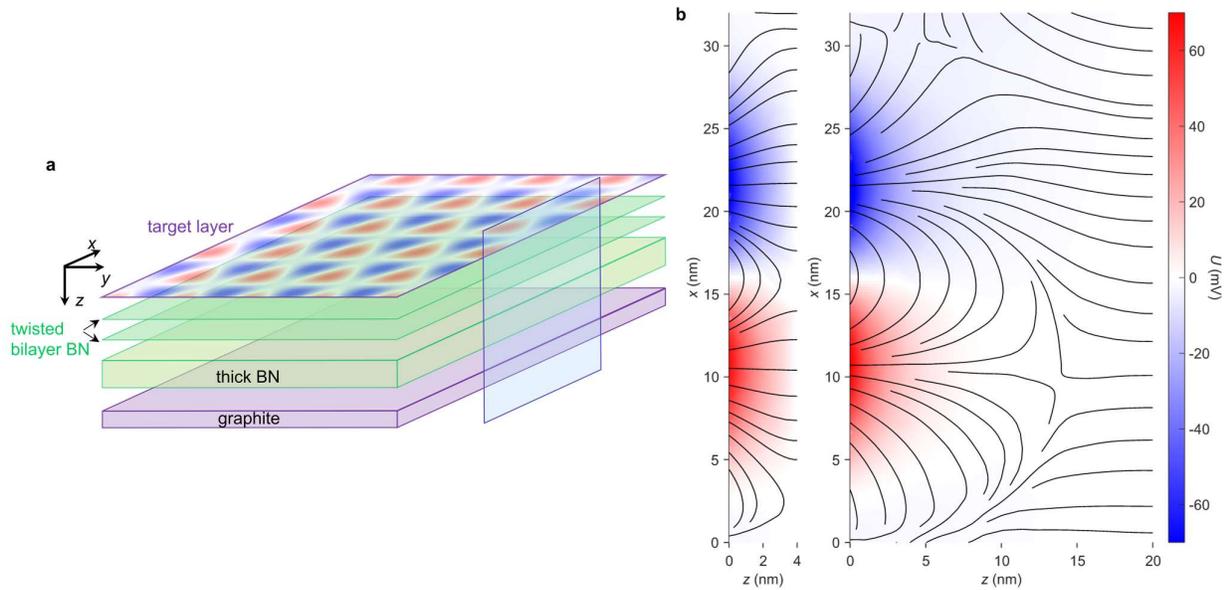

**Extended Data Fig. 5 | Electrostatic simulations of devices with different bottom BN thicknesses. a,** 3D view of the structure in simulation. The flakes lie in the *x-y* plane, and are stacked in *z* direction. The moiré pattern on top illustrates the electric potential on the target layer (bilayer graphene), beneath which the twisted bilayer BN, (bottom thick BN,) and bottom graphite are placed in order. The region shaded in blue denotes the orientation of cross-section view studied in **b.** The electric potential and the dimensions are not scaled in this schematic. **b,** Cross-section view of electric field lines and electric potential (*U*) in the device. Bottom BN thickness $d$ = 4.8 nm (left), and 20.8 nm (right). Compared with $d$ = 4.8 nm, more electric field lines are closed between adjacent AB and BA stackings in the twisted bilayer BN when $d$ = 20.8 nm. This explains why the electric potential profile becomes insensitive to the dielectric environment when the bottom BN thickness exceeds a certain threshold. Shown earlier in Fig. 3a, as *d* becomes larger, $U_{peak}$ remains almost constant.

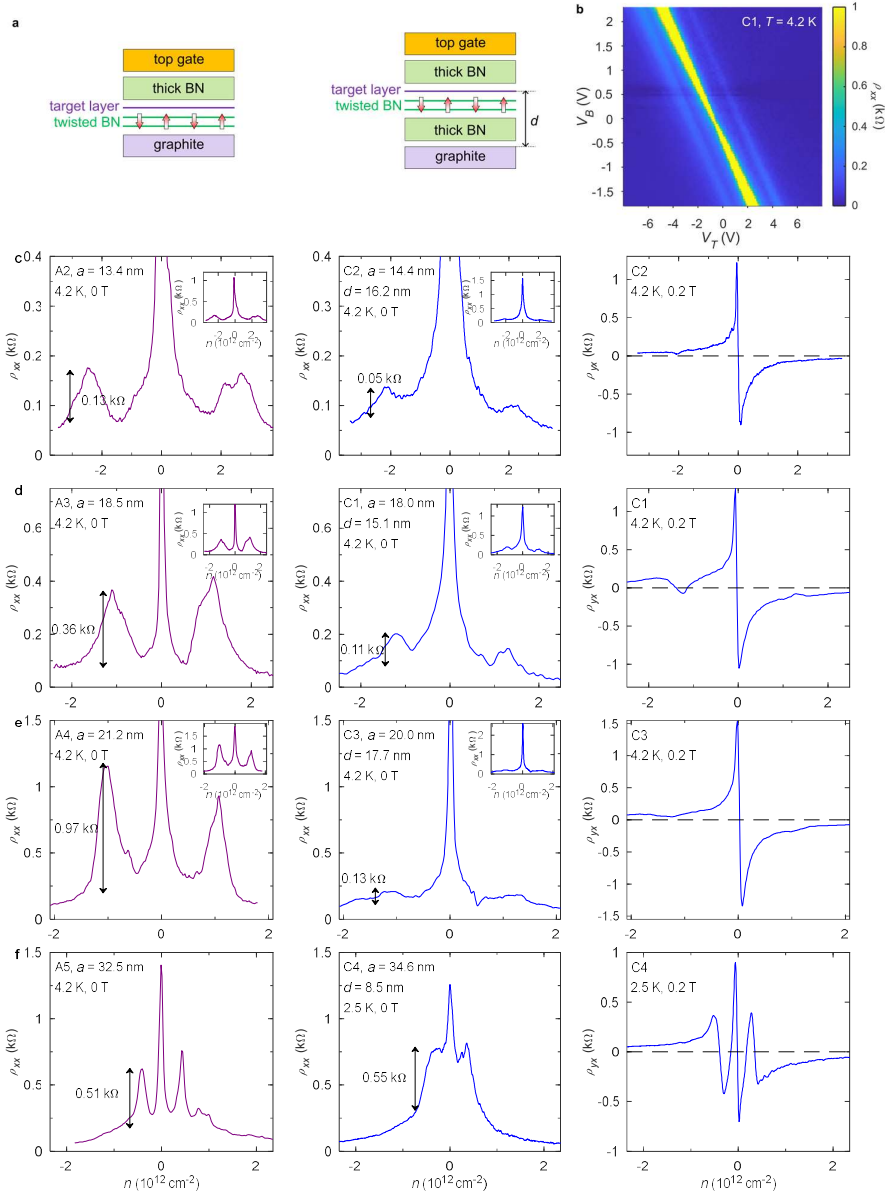

**Extended Data Fig. 6 | Comparison between devices with and without bottom thick BN. a,** Schematics of devices without (left) and with (right) bottom thick BN. **b,** Dual-gate scan of $\rho_{xx}$ in device C1 measured at 4.2 K. **c-f,** $\rho_{xx}$ as a function of $n$ in devices A2-A5 (left column, without bottom thick BN) and C1-C4 (middle column, with bottom thick BN). $\rho_{yx}$ as a function of $n$ in devices C1-C4, measured at $B = 0.2$ T (right column). Devices with similar moiré wavelengths are plotted side by side for comparison. The arrows indicate the height of satellite resistance peaks extracted from Lorentzian curve fitting. In most cases, the peak heights in devices without bottom thick BN are significantly higher than those with bottom BN. We notice that the peak heights of device A5 and C4 are comparable. We speculate the reason to be that the peak heights start to become less sensitive to the moiré potential strength when the moiré effect is strong enough, as they may start to be limited by twist angle disorders, etc. As explained in the main text, the moiré effect is affected by both the moiré wavelength which determines the kinetic energy, and the strength of moiré electrostatic potential. At large moiré wavelengths, the moiré effect is already strongly enhanced due to the its corresponding small kinetic energy. Therefore, the satellite resistance peak heights may become less sensitive to the moiré electrostatic potential strength. Insets: Large-scale plots.

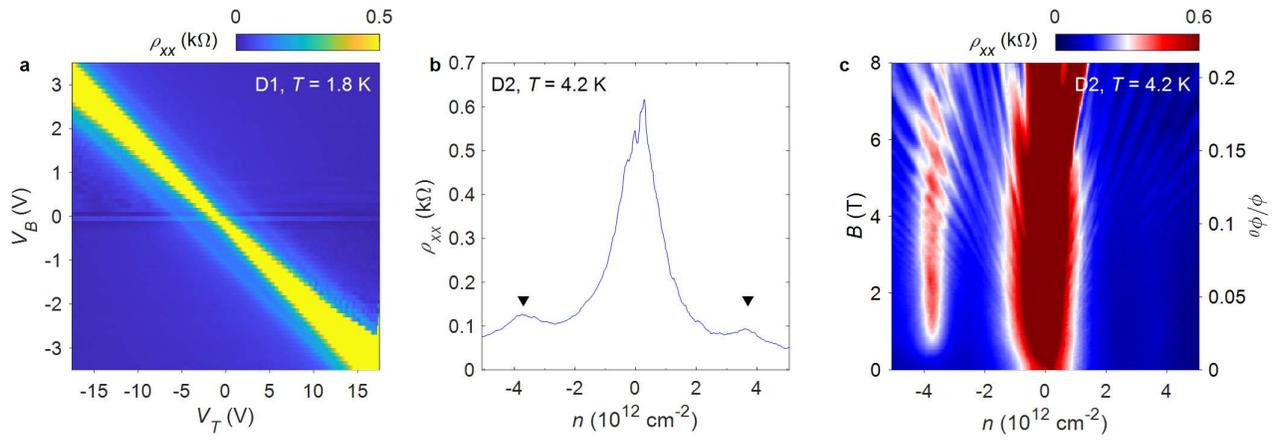

**Extended Data Fig. 7 | Additional data of devices with near-60°-twisted bilayer BN. a,** Dual-gate scan of $\rho_{xx}$ in device D1 measured at 1.8 K. **b,** $\rho_{xx}$ as a function of $n$ in device D2 measured at 4.2 K. **c,** $R_{xx}$ as a function of $n$ and $B$ (right $y$ axis: $\phi/\phi_0$) in device D2 at 4.2 K.